\documentclass[12pt,showpacs]{revtex4}
\usepackage{graphicx}
\begin{document}
\title{Analytic solutions for quantum logic gates and
modeling pulse errors in a quantum computer with
a Heisenberg interaction}
\author{G.P. Berman$^1$, D.I. Kamenev$^1$,
and V.I. Tsifrinovich$^2$}
\affiliation{$^1$Theoretical Division and Center for Nonlinear Studies,
Los Alamos National Laboratory, Los Alamos, New Mexico 87545}
\affiliation{$^2$IDS Department, Polytechnic University,
Six Metrotech Center, Brooklyn, New York 11201}
\vspace{3mm}
\begin{abstract}
We analyze analytically and numerically quantum logic 
gates in a one-dimensional spin chain with Heisenberg 
interaction. 
Analytic solutions for basic one-qubit gates
and swap gate are obtained for a quantum computer based on 
logical qubits. 
We calculated the errors caused by imperfect pulses which 
implement the quantum logic gates. 
It is numerically demonstrated that the probability error 
is proportional to $\varepsilon^4$, 
while the phase error is proportional to $\varepsilon$, 
where $\varepsilon$ is the characteristic deviation from  
the perfect pulse duration. 
\end{abstract}
\maketitle

\section{Introduction}
It is known that the Heisenberg interaction alone can provide
a universal set of gates for quantum computation \cite{DNature}. 
A computer based on the Heisenberg interaction
does not require magnetic fields nor
electromagnetic pulses. Implementations of a quantum computer 
using the Heisenberg interaction between 
the spins of the quantum dots or impurities 
in semiconductors promise clock-speeds in GHz region. 
The spins do not interact with each other unless one applies 
a voltage, which turns on the exchange interaction 
between a selected pair of spins.

In order to perform single-qubit rotations using the Heisenberg
interaction,
one should use coded, or logical, qubits. In this
paper, we use the coding introduced in Ref.~\cite{DNature} and
derive optimal gate sequences to implement swap 
gate and basic one-qubit logic operations. The errors caused
by imperfections of the pulses are investigated numerically.
The random deviations in the areas of the pulses in our 
simulations are assumed to have a Gaussian distribution 
with variance $\varepsilon$.

\section{Quantum dynamics}
Consider the dynamics of a spin system with an isotropic
Heisenberg interaction between neighboring spins.
The Hamiltonian which describes the interaction $J(\bar t)$
between $k$th and $(k+1)$th spins is
\begin{equation}
\label{Ht}
H_k(\bar t)=J(\bar t){\rm\bf S}_k{\rm\bf S}_{k+1},
\end{equation}
where $\bar t$ is time, ${\rm\bf S}_{k}$ is
the operator of the $k$th spin $1/2$.
The solution of the Schr\"odinger equation with the Hamiltonian
(\ref{Ht}) can be written in the form
\begin{equation}
\label{psiBart}
\psi(\bar t)=
\exp\left(-{i\over\hbar}\int_0^{\bar t}J(\bar t')d\bar t'
{\rm\bf S}_k{\rm\bf S}_{k+1}\right)\psi(0).
\end{equation}
(We do not use the time-ordering operator because
$[H_k(\bar t),H_k(\bar t^\prime)]=0$.)
Introducing a new effective dimensionless time,
\begin{equation}
\label{dimensionlesst}
t={1\over 2\pi\hbar}\int_0^{\bar t} J(\bar t')d\bar t',
\end{equation}
one can write Eq. (\ref{psiBart}) as
\begin{equation}
\label{psit}
\psi(t)=\exp\left(-i2\pi t
{\rm\bf S}_k{\rm\bf S}_{k+1}\right)\psi(0).
\end{equation}
This is the solution of the following dimensionless 
Schr\"odinger equation:
\begin{equation}
\label{Sch}
i{\partial\psi(t)\over\partial t}=V_k\psi(t),
\end{equation}
where
\begin{equation}
\label{Vk}
V_k=
2\pi\left[S_k^zS_{k+1}^z+
\frac 12(S_k^+S_{k+1}^-+S_k^-S_{k+1}^+)\right]
\end{equation}
is the dimensionless Hamiltonian,
$S_k^x$ $S_k^y$, and $S_k^z$ are the
components of the operator ${\rm\bf S}_{k}$,
$S_k^\pm=S_k^x\pm iS_k^y$.
After decomposition of the wave function in the
basis states $|n\rangle$,
\begin{equation}
\label{decomposition}
\psi(t)=\sum_nC_n(t)|n\rangle,
\end{equation}
where the states $|n\rangle$ are defined below
in Eqs. (\ref{physical}) and (\ref{statesSwap}),
one obtains a system of dimensionless differential
equations for the expansion coefficients,
\begin{equation}
\label{differential}
i\dot C_n=\sum_m\langle n|V_k|m\rangle C_m,
\end{equation}
where the dot indicates differentiation with respect to time $t$.

\begin{figure}
\centerline{\includegraphics[width=7cm,height=4cm]{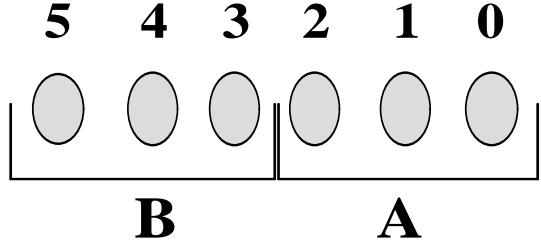}}
\vspace{-5mm}
\caption{Coding two logical qubits using six physical qubits.}
\label{fig:1}
\end{figure}

The matrix elements in Eq. (\ref{differential}) can be calculated
by action of the Hamiltonian $V_k$ in Eq. (\ref{Vk})
on the physical qubits (spins) using the following relations:
$$
S_k^z|\dots 0_k\dots\rangle=\frac 12|\dots 0_k\dots\rangle,\qquad
S_k^z|\dots 1_k\dots\rangle=-\frac 12|\dots 1_k\dots\rangle,
$$
\begin{equation}
\label{operators}
S_k^+|\dots 0_k\dots\rangle=0,~~\qquad
S_k^+|\dots 1_k\dots\rangle=|\dots 0_k\dots\rangle,
\end{equation}
$$
S_k^-|\dots 0_k\dots\rangle=|\dots 1_k\dots\rangle,\qquad
S_k^-|\dots 1_k\dots\rangle=0.~
$$

\section{Single qubit gates}
Let us consider only the first three spins, 0, 1, and 2, of the
spin chain in Fig. 1. We suppose that initially there 
are two spins in the state $|0\rangle$ and one spin 
in the state $|1\rangle$. 
Since the Hamiltonians $V_k$ can not flip individual spins 
(but can only swap the neighboring spins) 
one can choose an invariant subspace spanned by only three
states of the $2^3=8$ basis states:
\begin{equation}
\label{3states}
|0_20_11_0\rangle,~~~|0_21_10_0\rangle,~~~|1_20_10_0\rangle,
\end{equation}
or by their normalized and
orthogonal superpositions \cite{DNature},
$$
|0\rangle=|0_{\rm A}\rangle={1\over\sqrt 2}
(|0_21_10_0\rangle-|1_20_10_0\rangle),~~
|1\rangle=|1_{\rm A}\rangle=\sqrt{{2\over 3}}
\left(|0_20_11_0\rangle-{1\over 2}|0_21_10_0\rangle
-{1\over 2}|1_20_10_0\rangle\right),
$$
\begin{equation}
\label{physical}
|2\rangle=|a_{\rm A}\rangle={1\over \sqrt 3}
\left(|0_20_11_0\rangle+|0_21_10_0\rangle+
|1_20_10_0\rangle\right).
\end{equation}

We define the state $|0_{\rm A}\rangle$ as the ground state
of the logical qubit A; the state $|1_{\rm A}\rangle$
as the excited state; and the state $|a_{\rm A}\rangle$ as
the auxiliary state. One can show that all matrix elements
for transitions to the state $|a_{\rm A}\rangle$ are
equal to zero,
\begin{equation}
\label{me_a}
\langle a_A|V_0|0_A\rangle=\langle a_A|V_0|1_A\rangle=
\langle a_A|V_1|0_A\rangle=\langle a_A|V_1|1_A\rangle=0.
\end{equation}
If the state $|a_{\rm A}\rangle$ is initially not
populated, it remains empty under the action of the
Hamiltonians $V_0$ and $V_1$. For the single-qubit operations,
analyzed in this Section we
assume that initially $C_2(t=0)=0$ and we consider the dynamics
including only the states $|0\rangle=|0_A\rangle$ and
$|1\rangle=|1_A\rangle$.

The matrix elements of the two Hamiltonians have the form
\begin{equation}
\label{me}
V_0:~\pmatrix{
0 & -\Omega/2 \cr
-\Omega/2 & \Delta},\qquad
V_1:~\pmatrix{
3\Delta/2 & 0 \cr
0 & -\Delta/2},
\end{equation}
where
\begin{equation}
\label{frequencies}
\Delta=-\pi,\qquad \Omega=-\sqrt 3\pi.
\end{equation}

The solution of the Schr\"odinger equation generated by the
diagonal matrix $V_1$ has the form
\begin{equation}
\label{solution_d}
C_0(t)=e^{-i3\Delta t/2}C_0(0),\qquad
C_1(t)=e^{i\Delta t/2}C_1(0).
\end{equation}
The solution generated by the matrix $V_0$ is
$$
C_0(t)=\left\{C_0(0)\left[\cos(\Lambda t/2)+
i{\Delta\over\Lambda}\sin(\Lambda t/2)\right]+
iC_1(0){\Omega\over\Lambda}\sin(\Lambda t/2)\right\}
e^{-i\Delta t/2},
$$
\begin{equation}
\label{solution}
C_1(t)=\left\{C_1(0)\left[\cos(\Lambda t/2)-
i{\Delta\over\Lambda}\sin(\Lambda t/2)\right]+
iC_0(0){\Omega\over\Lambda}\sin(\Lambda t/2)\right\}
e^{-i\Delta t/2},
\end{equation}
where
\begin{equation}
\label{Lambda}
\Lambda=\sqrt{\Delta^2+\Omega^2}=2\pi.
\end{equation}
For convenience, we present below all dependences
expressed in terms of the frequencies $\Omega$, $\Delta$, and
$\Lambda$, but not in terms of their numerical values.

\subsection{One logical qubit flip}
In order to flip the logical qubit A using
Eq. (\ref{solution}) we assume
\begin{equation}
\label{initial1}
C_0(0)=1,~~ C_1(0)=0,
\end{equation}
and apply the Hamiltonian $V_0$ for time $t$.
Then, one obtains
$$
C_0(t)=\left[\cos(\Lambda t/2)+
i{\Delta\over\Lambda}\sin(\Lambda t/2)\right]e^{-i\Delta t/2},
$$
\begin{equation}
\label{flip1}
C_1(t)=i{\Omega\over\Lambda}\sin(\Lambda t/2)e^{-i\Delta t/2}.
\end{equation}
From this solution one can see that it is impossible to flip
the logical qubit using only one pulse since the coefficient
$C_0(t)$ in Eq. (\ref{flip1})
does not become zero for any $t$.
To solve this problem we use the
pulse sequence
\begin{equation}
\label{flipA}
{\rm F}_{\rm A}^{\rm ph}=V_0(t_3)V_1(t_2)V_0(t_1),
\end{equation}
proposed in Ref. \cite{DNature}. 
Here and below the superscript
`ph' indicates that the gate requires additional pulses
to implement the phase correction.
In Eq. (\ref{flipA}) $V_i(t)$ indicates action of
$i$th Hamiltonian during time $t$, and the sequence
must be read
from right to left. 
In this Section we obtain exact analytical
expressions for $t_1$, $t_2$, and $t_3$.

A flip of the qubit A with the initial conditions (\ref{initial1})
means making the transition $|0\rangle\rightarrow|1\rangle$.
Using Eqs. (\ref{solution_d}), (\ref{solution}), and
(\ref{initial1}) and setting the
amplitude $C_0(t)=0$ after the action
of the ${\rm F}^{\rm ph}_{\rm A}$ gate, one obtains the equation
$$
e^{-i{\Delta\over 2}(t_1-t_2+t_3)}
\left\{e^{-2i\Delta t_2}\left[\cos(\Lambda t_1/2)+
i{\Delta\over\Lambda}\sin(\Lambda t_1/2)\right]
\left[\cos(\Lambda t_3/2)+
i{\Delta\over\Lambda}\sin(\Lambda t_3/2)\right]-\right.
$$
\begin{equation}
\label{flip2}
\left.
{\Omega^2\over\Lambda^2}\sin(\Lambda t_1/2)\sin(\Lambda t_3/2)
\right\}=0.
\end{equation}
Equation (\ref{flip2}) is satisfied when
both the real and the imaginary parts
of the expression in the curly brackets are equal to zero.

In order to solve Eq. (\ref{flip2}) we first assume that
$\cos(\Lambda t_1/2)\ne 0$, $\cos(2\Delta t_2)\ne 0$, and
$\cos(\Lambda t_3/2)\ne 0$. Then, for
$$
x=\tan(\Lambda t_1/2),~~y=\tan(2\Delta t_2),
~~z=\tan(\Lambda t_3/2)
$$
one obtains the following system of two coupled equations:
$$
1-xz+{\Delta\over\Lambda}y(x+z)=0,
$$
\begin{equation}
\label{flip3}
y\left(1-{\Delta^2\over\Lambda^2}xz\right)+
{\Delta\over\Lambda}(x+z)=0.
\end{equation}
Using Eqs. (\ref{frequencies}) and (\ref{Lambda}) 
and eliminating $y$, one has
\begin{equation}
\label{flip4}
(xz-1)(4-xz)=(x+z)^2.
\end{equation}
Introducing the notations $x+z=2b$ and $xz=c$ one can present
$x$ and $z$ as the two solutions ($\xi_1=x$ and $\xi_2=z$)
of the quadratic equation
\begin{equation}
\label{flip5}
\xi^2-2b\xi+c=0.
\end{equation}
Using definitions of $b$ and $c$ and Eq. (\ref{flip4})
one can show
that Eq. (\ref{flip5}) has no real solution.

In a similar way one can show that there is no real solution when
$\cos(\Lambda t_1/2)=0$ or $\cos(\Lambda t_3/2)=0$.
For $\cos(2\Delta t_2)=0$, the two solutions are
$$
{t'}_1={{\rm arctan}(3-\sqrt 5)\over \pi},~~{t'}_2=\frac 14,~~
{t'}_3={{\rm arctan}(3+\sqrt 5)\over \pi}
$$
and
\begin{equation}
\label{flip6}
t_1=1-{{\rm arctan}(3-\sqrt 5)\over \pi},~~t_2=\frac 34,~~
t_3=1-{{\rm arctan}(3+\sqrt 5)\over \pi}.
\end{equation}
Below we use only the second solution (\ref{flip6}).

The ${\rm F}_{\rm A}^{\rm ph}$ gate generates
different phases for two basis states,
\begin{equation}
\label{flip7}
{\rm F}_{\rm A}^{\rm ph}\pmatrix{C_0 \cr C_1}=
\pmatrix{e^{i\varphi_2}C_1 \cr e^{i\varphi_1}C_0},
\end{equation}
where
\begin{equation}
\label{phase1}
\varphi_1=\frac 12\left[\frac{3\pi} 4+{\rm arctan}~2-
{\rm arctan}(\sqrt 5/2)\right],~~~
\varphi_2=\frac 12\left[\frac{3\pi} 4+{\rm arctan}~2+
{\rm arctan}(\sqrt 5/2)\right].
\end{equation}
In order to correct the phases, an additional pulse is required.
The phase-corrected gate ${\rm F}_{\rm A}$ for flipping
the qubit A has the form
\begin{equation}
\label{flip8}
{\rm F}_{\rm A}=V_1(t_4)F_{\rm A}^{\rm ph}=
V_1(t_4)V_0(t_3)V_1(t_2)V_0(t_1).
\end{equation}
In order to find the time $t_4$ we use the
solution (\ref{solution_d}).
The additional phase-correcting pulse $V_1(t_4)$
modifies Eq. (\ref{flip7}) to become
\begin{equation}
\label{flip9}
{\rm F}_{\rm A}\pmatrix{C_0 \cr C_1}=
\pmatrix{e^{i(\varphi_2-3\Delta t_4/2)}C_1
\cr e^{i(\varphi_1+\Delta t_4/2)}C_0}.
\end{equation}
One can make the phases of the both states equal to each other by
application of the ${\rm F}_{\rm A}$ gate if the condition
$$
\varphi_1+\Delta t_4/2=\varphi_2-3\Delta t_4/2
$$
is satisfied. This equation determines the last parameter,
\begin{equation}
\label{t4}
t_4=1-{{\rm arctan(\sqrt 5/2)}\over 2\pi},
\end{equation}
required to implement the phase-corrected
flip of the qubit A. The flip
gate for the qubit A can be written as
\begin{equation}
\label{flips}
{\rm F}_{\rm A}\pmatrix{C_0 \cr C_1}=
e^{i\Phi_{\rm F}}\pmatrix{C_1 \cr C_0},
\end{equation}
where the overall phase for the single qubit flip gate is
\begin{equation}
\label{phaseF}
\Phi_{\rm F}=-{\pi\over 8}+{1\over 2}\,{\rm arctan}~2-
{1\over 4}\,{\rm arctan(\sqrt 5/2)}.
\end{equation}

\subsection{Hadamard transform}
The Hadamard transform H$_{\rm A}$ for the qubit A can be
performed using the pulse sequence
\begin{equation}
\label{Hadamard}
{\rm H}_{\rm A}=V_1(t_5)V_0(t_6)V_1(t_5).
\end{equation}
Here the pulses $V_1(t_5)$ are used to provide the correct
phases and the pulse $V_0(t_6)$ is needed to split each of the
states $|0\rangle$ and $|1\rangle$
into a superposition of the states with equal probabilities.
The time-intervals are
\begin{equation}
\label{time_H}
t_5=\frac 34+{{\rm arctan}(1/\sqrt 2)\over 2\pi},\qquad
t_6={{\rm arctan}\sqrt 2\over \pi}.
\end{equation}
The Hadamard gate transforms the
wave function as
\begin{equation}
\label{Hadamards}
{\rm H}_{\rm A}\pmatrix{C_0 \cr C_1}=
{e^{i\pi/2}\over\sqrt 2}
\pmatrix{C_0+C_1 \cr C_0-C_1}.
\end{equation}

\subsection{Phase gate}
The phase gate P$_{\rm A}(\theta)$ for the qubit A
can be performed using only one pulse
\begin{equation}
\label{Phase}
{\rm P}_{\rm A}(\theta)=V_1[t(\theta)],
\end{equation}
where
\begin{equation}
\label{tTheta}
t(\theta)=1-{\theta\over 2\pi},
\end{equation}
and the angle $\theta$ is defined in the interval $[0,2\pi]$.
The Phase gate transforms the
wave function in the following way
\begin{equation}
\label{phasess}
{\rm P}_{\rm A}\pmatrix{C_0 \cr C_1}=
e^{i\Phi_{\rm P}(\theta)}
\pmatrix{C_0 \cr e^{i\theta}C_1}.
\end{equation}
The overall phase generated by the phase gate is
\begin{equation}
\label{phasePhase}
\Phi_{\rm P}(\theta)={3\pi\over 2}-{3\theta\over 4}.
\end{equation}

The single qubit operations for the qubit B in Fig. 1
can be performed using the same sequences
like those for the qubit A with the
substitutions $V_0(t)\rightarrow V_3(t)$ and
$V_1(t)\rightarrow V_4(t)$.

\section{Swap gate}
It is convenient to
analyze the spin states [from which the
logical qubits are formed, see Eq. (\ref{physical})].
Consider the four different spin states,
$$
|m\rangle=|\dots 0_k0_{k+1}\dots\rangle,~~~
|p\rangle=|\dots 1_k1_{k+1}\dots\rangle,
$$
\begin{equation}
\label{4states}
|i\rangle=|\dots 1_k0_{k+1}\dots\rangle,~~~
|j\rangle=|\dots 0_k1_{k+1}\dots\rangle.
\end{equation}
These states form two one-dimensional and one 
two-dimensional invariant subspaces. The states
$|m\rangle$ and $|p\rangle$ are eigenstates of
the Hamiltonian $V_k$,
\begin{equation}
\label{mp}
V_k(t)|m\rangle=e^{-i\frac\pi 2t}|m\rangle,~~~
V_k(t)|p\rangle=e^{-i\frac\pi 2t}|p\rangle.
\end{equation}
The states $|i\rangle$ and $|j\rangle$ are transformed as
$$
V_k(t)|i\rangle=e^{i\frac\pi 2t}\left[
\cos(\pi t)|i\rangle-i\sin(\pi t)|j\rangle\right],
$$
\begin{equation}
\label{ij1}
V_k(t)|j\rangle=e^{i\frac\pi 2t}\left[
\cos(\pi t)|j\rangle-i\sin(\pi t)|i\rangle\right].
\end{equation}
From Eqs. (\ref{mp}) and (\ref{ij1}) one can see that the pulse
$V_k(1/2)$ can be used as a swap gate between the
$k$th and $(k+1)$th spins.
After the pulse $V_k(1/2)$ all states acquire the phase
$-i\pi/4$.

The swap gate between the spins can be used for 
implementation of the swap gate
between the logical qubits. 
The two logical qubits, A and B, in Fig. 1 are formed by
the superpositions of the spin states involving six spins.
Consider one state of the superposition, 
for example, the state $|0_50_41_3~0_21_10_0\rangle$.
The spins 0, 1, and 2 are related to the logical qubit A and
the spins 3, 4, and 5 are related to the logical qubit B. 
Using five swaps between the neighboring spins one can 
move the state of the 5th spin to the zeroth spin,
\begin{equation}
\label{cyclic}
C_{5,0}|{\bf 0_5}0_41_3~0_21_10_0\rangle=e^{-i{5\pi\over 4}}
|0_51_40_3~1_20_1{\bf 0_0}\rangle,
\end{equation}
where ${\rm C}_{5,0}$ is the operator or the cyclic permutation,
\begin{equation}
\label{C}
C_{5,0}=V_0\left(\frac 12\right)
V_1\left(\frac 12\right)V_2\left(\frac 12\right)
V_3\left(\frac 12\right)V_4\left(\frac 12\right).
\end{equation}
Three successive applications of the operator ${\rm C}_{5,0}$
result in the swap gate S$_{\rm AB}$ between the logical qubits
(below called swap gate),
\begin{equation}
\label{Swap}
{\rm S}_{\rm AB}=C_{5,0}C_{5,0}C_{5,0}.
\end{equation}
The swap gate $S_{\rm AB}$ produces
an overall phase $\pi/4$ for the wave function. 
The total number of pulses required to execute the swap gate is 15.
Note that the result of the swap gate is independent
of a kind of coding of the logical qubits through the spin states.

\section{Modeling errors in the swap gate}
In spite of the rather simple form of the swap gate 
S$_{\rm AB}$,
it does implement a complex logic operation on logical qubits.
Indeed, if initially one has a basis logical state,
e.g. $|1_{\rm B}0_{\rm A}\rangle$, 
in the process of applying the swap gate one has a 
superposition of many states, while, finally, only one 
state ($|0_{\rm B}1_{\rm A}\rangle)$ survives, and 
all other states disappear.  

Numerical simulations of the swap gate between the
qubits A and B were performed in the invariant 
Hilbert subspace spanned by
the following 15 [$15=C^2_6={6!/(2!4!)}$] states:
$$
|0\rangle=|0_{\rm B}0_{\rm A}\rangle,~~
|1\rangle=|0_{\rm B}1_{\rm A}\rangle,~~
|2\rangle=|1_{\rm B}0_{\rm A}\rangle,~~
|3\rangle=|1_{\rm B}1_{\rm A}\rangle,~~
$$
$$
|4\rangle=|0_{\rm B}a_{\rm A}\rangle,~~
|5\rangle=|1_{\rm B}a_{\rm A}\rangle,~~
|6\rangle=|a_{\rm B}0_{\rm A}\rangle,~~
|7\rangle=|a_{\rm B}1_{\rm A}\rangle,~~
$$
\begin{equation}
\label{statesSwap}
|8\rangle=|a_{\rm B}a_{\rm A}\rangle,~~
|9\rangle=|000~011\rangle,~~
|10\rangle=|000~101\rangle,~~
|11\rangle=|000~110\rangle,~~
\end{equation}
$$
|12\rangle=|011~000\rangle,~~
|13\rangle=|101~000\rangle,~~
|14\rangle=|110~000\rangle.
$$
The dynamics was simulated using the
evolution operators built using the eigenstates of the
Hamiltonians $V_k$, $k=0,1,\dots,5$, in the 15-dimensional space.
When the time-intervals $t$ for the 
pulses were exactly
equal to $t=t_0=1/2$, the errors 
in implementation of the swap gate
were of the order of $10^{-15}$, i.e. accuracy was limited only 
by the round-off errors. Since $t$ is proportional to the
area of the pulse, the form of the pulse is not important. 
However, in an experiment there is always some deviation in $t$ 
from its optimal value $t_0$. 
To understand the error caused by this deviation,  
we modeled the swap gate with imperfect pulses. 
The duration of each imperfect pulse is taken as
\begin{equation}
\label{tgauss}
t=t_0+\delta t,
\end{equation}
where the random deviation $\delta t$ is assumed to have the
Gaussian distribution $\exp[-(\delta t)^2/(2\varepsilon^2)]$.

\begin{figure}
\mbox{
\includegraphics[width=8cm,height=7cm]{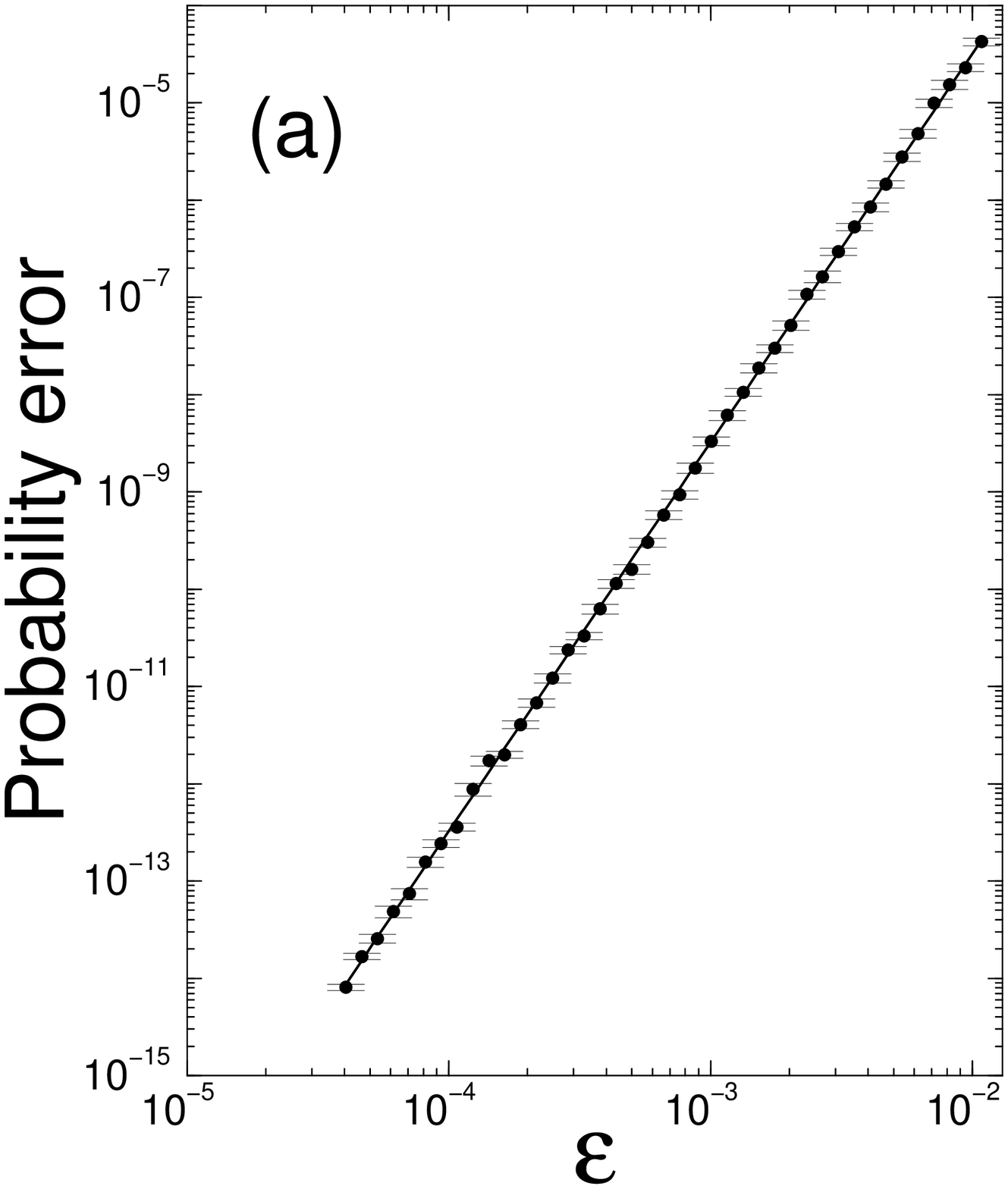}
\includegraphics[width=8cm,height=7cm]{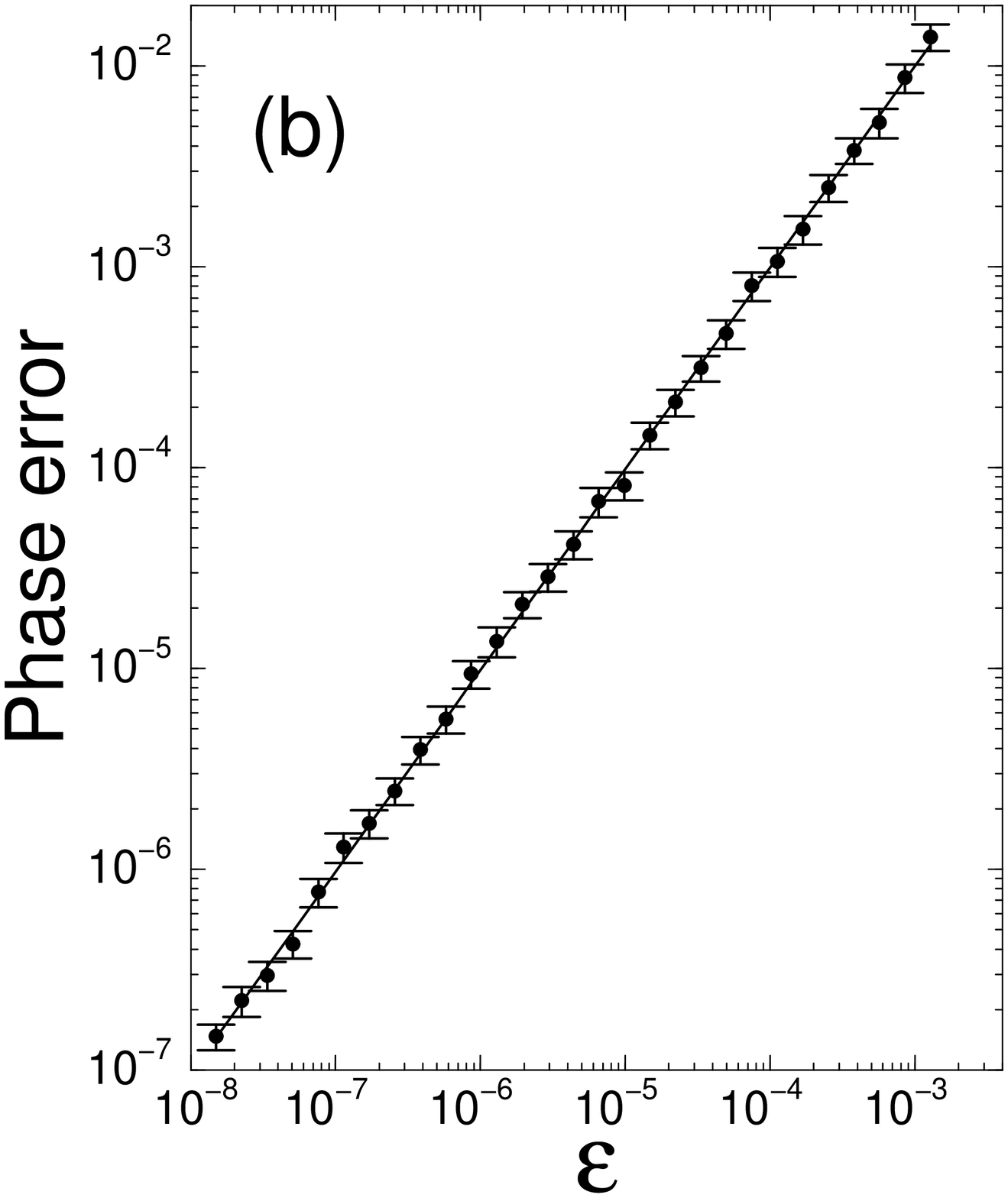}}
\vspace{-5mm}
\caption{(a) The average probability error $P_{\rm S}$
and (b) the average phase error $Q_{\rm S}$
of the swap gate 
as a function of 
$\varepsilon$ (filled circles). The least square fits 
(solid lines), show that: 
(a) the probability error increases as 
$P_{\rm S}= 
3.183\times 10^3\times\varepsilon^{3.998}$, $\chi^2=43.5$; 
(b) the phase error is given by
$Q_{\rm S}= 10.2033\times\varepsilon^{1.0031}$, 
$\chi^2=1.0$.} 
\label{fig:2}
\end{figure}

We define the probability error as 
\begin{equation}
\label{eprobability}
P_{\rm S}=||C_j(T)|^2-|C_i(0)|^2|,~~~C_i(0)=1,
\end{equation}
where $T$ is the duration of the swap gate and 
the final state $|j\rangle$ is related to the initial state 
$|i\rangle$ as 
\begin{equation}
\label{ij}
|j\rangle={\rm S}^{\rm i}_{\rm AB}|i\rangle.
\end{equation}
Here ${\rm S}^{\rm i}_{\rm AB}$ is the ideal swap gate. 
The probability error $P_{\rm S}$, shown in 
Fig. 2(a), increases 
as a function of $\varepsilon$ approximately 
as $3.2\times 10^3\times\varepsilon^4$.  

Next, we study the phase errors [see Fig. 2(b)], caused by the 
random fluctuation of the pulse duration $t$. Under the 
action of the sequence (\ref{Swap}) of the
perfect pulses the four logical basis states 
$|00\rangle$, $|01\rangle$, $|10\rangle$, and $|11\rangle$
transform correspondingly to 
$|00\rangle$, $|10\rangle$, $|01\rangle$, and $|11\rangle$ 
with the same phase shift. Under the action of the imperfect 
pulses we obtain four different phase shifts for the basis 
states. We define the phase error $Q_{\rm S}$
as the maximum difference between these phase shifts.
From Fig. 2(b) one can see that the phase error is 
approximately equal to $10.2\varepsilon$. 

The data in Figs.~2(a,b) are averaged over 1000 runs with 
different randomly chosen initial states $|i\rangle$ and 
different random deviations $\delta t$ from the ideal pulse 
duration $t_0$. 
In Figs. 2(a,b) $\chi^2$ was calculated as~\cite{chi2}
$$
\chi^2=\sum_{i=1}^K{(y_k-\bar y_k)^2\over(\delta y_k)^2}, 
$$
where the index $k$ labels the points on the graphs, 
$K$ is the number of points, $y_k=P_{\rm S}^k$ 
in Fig. 2(a)
and $y_k=Q_{\rm S}^k$ in Fig. 2(b) are the 
coordinates of the circles, $\bar y_k$ are 
the corresponding coordinates of the points on the straight 
lines for the same values of $\varepsilon$; 
$\delta y_k$ are the corresponding standard deviations. 

\section{Conclusion}
In this paper, analytic solutions for quantum logic
gates are obtained
for a quantum computer with an isotropic Heisenberg interaction 
between neighboring identical spins arranged in a 
one-dimensional spin chain. 
Single qubit flip, Hadamard and phase transforms
are implemented by using, respectively,
4, 3, and 1 pulse(s). The swap gate is realized using 15 pulses.
The probability and phase errors caused by 
imperfect pulses for the swap gate are calculated numerically.
The probability error
is proportional to $\varepsilon^4$, 
while the phase error is proportional to $\varepsilon$, 
where $\varepsilon$ is the characteristic
deviation from the perfect pulse duration.

\begin{acknowledgments}
We thank G. D. Doolen for useful discussions.
This work was supported by the Department of Energy (DOE) under
Contract No. W-7405-ENG-36, by the National Security Agency (NSA),
and by the Advanced Research and Development Activity (ARDA).
\end{acknowledgments}

{}
\end{document}